\documentclass[conference,10pt]{IEEEtran}
\IEEEoverridecommandlockouts
\usepackage{cite}
\usepackage{amsmath,amssymb,amsfonts}
\usepackage{algorithmic}
\usepackage{graphicx}
\usepackage{textcomp}
\usepackage{xcolor}

\usepackage{amsthm,mathtools}
\mathtoolsset{showonlyrefs=true} 
\usepackage{bbm}

\mathchardef\Re="023C
\mathchardef\Im="023D
\usepackage{breqn}
\usepackage{float}
\usepackage{caption}
\usepackage{subcaption}
\usepackage{gensymb}
\usepackage{verbatim}
\usepackage{euscript}

\usepackage{standalone}

\DeclareMathOperator{\E}{\mathbb{E}}

\newcommand{\rem}[1]{}

\newcommand{\bre}{\begin{equation}}
\newcommand{\ere}{\end{equation}}

\newcommand{\ee}\]
\newcommand{\bfg}{\begin{figure}[hbtp]}
\newcommand{\efg}{\end{figure}}
\newcommand{\bver}{\begin{verbatim}}
\newcommand{\ever}{\end{verbatim}}
\newcommand{\bit}{\begin{itemize}}
\newcommand{\eit}{\end{itemize}}
\newcommand{\ben}{\begin{enumerate}}
\newcommand{\een}{\end{enumerate}}

\newcommand{\bsigma}{\boldsymbol\sigma}

\newcommand{\btheta}{\boldsymbol\theta}

\newcommand{\bmu}{\boldsymbol\mu}

\newcommand{\given}{\: | \:}

\newcommand{\bphi}{{\mathbf \Phi}}

\newcommand{\baa}{\begin{eqnarray*}}
\newcommand{\eaa}{\end{eqnarray*}}

\newcommand{\bs}{{\bf s}}

\newcommand{\bx}{{\bf x}}
\newcommand{\by}{{\bf y}}
\newcommand{\bz}{{\bf z}}

\newcommand{\cL}{{\cal L}}

\newcommand{\cH}{{\cal H}}

\newcommand{\cC}{{\mathcal{C}}}

\newcommand{\cM}{{\mathcal{M}}}
\newcommand{\cN}{{\mathcal{N}}}

\newcommand{\defined}{\triangleq}

\def\argmax{\mathop{\rm argmax}}

\def\defined{\: {\stackrel{\scriptscriptstyle \Delta}{=}} \: }

\newfont{\boldlarge}{msbm10 scaled 1100}

\newlength{\tmpbigbar}

\def\BibTeX{{\rm B\kern-.05em{\sc i\kern-.025em b}\kern-.08em
    T\kern-.1667em\lower.7ex\hbox{E}\kern-.125emX}}

\begin{document}

\title{Semi-Supervised Variational Inference over Nonlinear Channels\\
\thanks{This research was supported by the Israel Science Foundation, grant no. 1868/18.}
}

\author{\IEEEauthorblockN{David~Burshtein}
\IEEEauthorblockA{\textit{School of Electrical Engineering} \\
\textit{Tel Aviv University}\\
Tel Aviv 6997801, Israel\\
burstyn@eng.tau.ac.il}
\and
\IEEEauthorblockN{Eli Bery}
\IEEEauthorblockA{\textit{School of Electrical Engineering} \\
\textit{Tel Aviv University}\\
Tel Aviv 6997801, Israel \\
bery2590@gmail.com}
}

\maketitle

\begin{abstract}
Deep learning methods for communications over unknown nonlinear channels have attracted considerable interest recently. In this paper, we consider semi-supervised learning methods, which are based on variational inference, for decoding unknown nonlinear channels. These methods, which include Monte Carlo expectation maximization and a variational autoencoder,  make efficient use of few pilot symbols and the payload data. The best semi-supervised learning results are achieved with a variational autoencoder. For sufficiently many payload symbols, the variational autoencoder also has lower error rate compared to meta learning that uses the pilot data of the present as well as previous transmission blocks.
\end{abstract}


\section{Introduction}
Deep learning methods for communications over unknown channels have attracted considerable interest recently. One of the challenges concerns the efficient use of pilot data \cite{park2020learning, shlezinger2020viterbinet} in order to decode reliably without a significant rate loss.
When transmitting over a linear intersymbol interference (ISI) channel, self-supervised decision directed methods can be used for tracking moderate changes in the channel, but typically they cannot be used for new channel identification.
Instead, unsupervised (blind) channel estimation over a linear ISI channel, up to a delay and rotation of the constellation, can be achieved with the constant modulus algorithm (CMA) \cite{godard1980self,treichler1983new}.
The maximum likelihood criterion can be used for improved blind channel equalization. However, due to computational intractability, rather than maximizing the likelihood function, the common practice is to use the expectation-maximization (EM) algorithm or an approximated version thereof \cite{kaleh1994joint, tong1998multichannel}.

Recently, variational autoencoders (VAEs) \cite{kingma2013auto,rezende2014stochastic} were proposed for blind channel equalization \cite{caciularu2020unsupervised}.
VAEs were shown to compare favorably with other methods for blind channel equalization over unknown linear channels, such as CMA or EM \cite{caciularu2020unsupervised}, in terms of the error rate obtained or the computational efficiency of the method. However, blind VAE equalizers still do not perform well for unknown non-linear channels, even if coding information is incorporated.
Hence, as an alternative to unsupervised blind variational inference, in this paper we propose semi-supervised variational inference, following the success of these methods in computer vision problems \cite{kingma2014semi}.
The loss function used for semi-supervised learning (SSL) involves both the labeled (pilot) symbols and the unlabeled (payload) symbols. It generalizes the loss used for unsupervised blind channel estimation \cite{caciularu2020unsupervised}.

In this paper we derive variational inference-based SSL methods that make efficient use of few pilots and the payload data when transmitting over unknown nonlinear channels.
We use the same communication example as in \cite{park2020learning, cohen2021learning}.
The best SSL results are achieved with a VAE. For sufficiently many payload symbols, the VAE also has lower error rate compared to meta learning \cite{park2020learning} that uses the pilot data of the present as well as previous transmission blocks.

\section{Problem formulation} \label{sec:problem_setup}
Consider a block of symbols, $(s_1,\ldots,s_N)$, transmitted over a memoryless channel $p(\by| s)$. Each symbol, $s_i \in \left\{ 1,2,\ldots,K \right\}$, determines a transmitted signal, $\bx_i=\bx(s_i)$, from a fixed constellation $\bx(\cdot)$ of size $K$. As in \cite{park2020learning, cohen2021learning}, we assume that the symbols are statistically independent uniformly distributed, $P(S_i=s)=1/K$, $s=1,2,\ldots,K$, transmitted over a memoryless channel. We denote by $(\by_1,\ldots,\by_N)$ the channel output, and by $p(\by_i| s_i)$ the conditional probability density function (PDF) of $\by_i$ given $s_i$.
The channel is unknown at the receiver. However, $N_p$ symbols, e.g. $\left\{ s_i \right\}_{i=1}^{N_p}$, are known pilot symbols. Conventional learning methods use the given pilot data information $\left\{ s_i, \by_i \right\}_{i=1}^{N_p}$ to train a decoder. For example, we can train a neural network decoder with input, $\by$, and a softmax output layer that produces a probability vector of size $K$, denoted by $\left\{ q_\bphi(s| \by_i) \right\}_{s=1}^K$, where $\bphi$ is the neural network parameters vector. Typically, we train the network by minimizing the cross entropy loss function,
\begin{equation}
	-\frac{1}{N_p}\sum_{i=1}^{N_p} \log q_\bphi (s_i | \by_i)
	\label{eq:CE_labeled}
\end{equation}
with respect to $\bphi$ using gradient descent (GD). The trained distribution, $\left\{ q_\bphi(s| \by) \right\}_{s=1}^K$, can be viewed as an approximation to the true posterior, $P(s| \by)$.
Denoting by $\hat{\bphi}$ the value of $\bphi$ obtained by the GD training, we decode the payload data, $\hat{s}_i$, $i=N_p+1,\ldots,N$, using
\begin{equation}
	\label{eq:decoder1}
	\hat{s}_i = \argmax_{s\in\{1,2,\ldots,K\}} q_{\hat{\bphi}}(s\given \by_i)
\end{equation}

\section{SSL for channel decoding} \label{sec:semisup}
To reduce the number of required pilot symbols, we consider SSL methods, that use the payload data, $\left\{ \by_i \right\}_{i=N_p+1}^{N}$, in addition to the pilot symbols, $\left\{ s_i, \by_i \right\}_{i=1}^{N_p}$.

\subsection{Simple Decision directed (SDD)}
To apply simple decision directed (SDD) SSL, we first obtain an initial estimate to $\bphi$, denoted $\bphi_0$, by minimizing the loss \eqref{eq:CE_labeled}, associated with the pilot (labeled) data. We use $\bphi_0$ to label the payload data, i.e.,
$\hat{s}_i = \argmax_{s=1,2,\ldots,K} q_{\bphi_0}(s | \by_i)$, $i=N_p+1,\ldots,N$.
Then, we minimize a second loss function,
\begin{equation}
	-\frac{\gamma_0}{N_p}\sum_{i=1}^{N_p} \log q_\bphi (s_i | \by_i)
	-\frac{1-\gamma_0}{N-N_p}\sum_{i=N_p+1}^{N} \log q_\bphi (\hat{s}_i | \by_i) 
	\label{eq:CE_DD}
\end{equation}
with respect to $\bphi$ ($\gamma_0\in(0,1)$ is a hyper-parameter). Denoting the minimizing value by $\hat{\bphi}$, we use \eqref{eq:decoder1} to decode the data.

\subsection{Monte Carlo EM (MCEM) and Viterbi EM} \label{sec:MCEM}
We suggest the following SSL method based on the EM algorithm \cite{dempster1977maximum}. We first consider a loss function, based on a weighted ML criterion, applied to all the available data, $\left\{ s_i,\by_i \right\}_{i=1}^{N_p}$ and $\left\{ \by_i \right\}_{i=N_p+1}^N$ ($\gamma\in(0,1)$ is a hyper-parameter),
\begin{equation}
\cL_{\text{ML}} =
-\frac{\gamma}{N_p} \sum_{i=1}^{N_p} \log p_\btheta(s_i, \by_i)
-\frac{1-\gamma}{N-N_p}\sum_{i=N_p+1}^{N} \log p_\btheta(\by_i)
\label{eq:MLLoss}
\end{equation}
\begin{equation}
p_{\btheta}(s,\by) = p(s) p_{\btheta}(\by | s) = \frac{1}{K} p_{\btheta}(\by | s)
\label{eq:p_sy}
\end{equation}
where $p_{\btheta}(\by | s)$ is a model, parameterized by $\btheta$, for the conditional PDF of $\by$ given $s$. In our experiments we model $p_{\btheta}(\by | s)$ by a diagonal covariance Gaussian PDF, $\cN(\by;\bmu_\btheta \left( \bx(s)), \bsigma_\btheta^2(\bx(s)) \right)$, so that
\begin{align}
\lefteqn{\log p_{\btheta}(\by | s) = -\frac{d}{2} \log (2\pi)}\\
&&-\frac{1}{2} \sum_{j=1}^{d} 
\left( \frac{(y_j-\mu_{\btheta,j}(\bx(s)))^2}{\sigma_{\btheta,j}^2(\bx(s))} + \log \sigma_{\btheta,j}^2(\bx(s)) \right) 
\label{eq:log_pys}
\end{align}
where $d$ is the dimension of $\by$, and where a neural network with input $\bx=\bx(s)$, and parameters vector, $\btheta$, produces the mean vector and the vector of log-variances, $\left( \bmu_\btheta(\bx), \log \bsigma_\btheta^2(\bx) \right)$, at its output.
Also,
$p_\btheta(\by_i) = \sum_s p_{\btheta}(s,\by_i) = \frac{1}{K} \sum_s p_{\btheta}(\by_i | s)$.
The loss function \eqref{eq:MLLoss} assigns different weighting to labeled and non-labeled data. To minimize this loss we can use GD. However, computing the gradient of this loss function requires application of the back-propagation algorithm $K$ times for each payload data measurement, $\by_i$. This may be computationally prohibitive for large values of $K$. Furthermore, this approach is computationally intractable for the case of channels with memory. As an alternative, we may use the following well known \cite{bishop2006pattern,kingma2013auto} evidence lower bound (ELBO),
\begin{equation}
	\log p_{\btheta} (\by) \geq \E_{q(s|\by)}
	\left[-\log q(s|\by)+\log p_{\btheta} (s,\by)\right]
	\label{eq:elbo}
\end{equation}
which is variational lower bound that holds for any conditional distribution, $q(s|\by)$. It holds with equality for $q(s|\by) = p_{\btheta} (s|\by)$ \cite{bishop2006pattern}.
The EM algorithm \cite{dempster1977maximum} is an iterative algorithm for computing the ML estimate. It can be derived from the ELBO \cite{bishop2006pattern}, by minimizing the upper bound on the log-likelihood repeatedly, first with respect to $q(\cdot|\cdot)$ while $\btheta$ is fixed, and then with respect to $\btheta$ while $q(\cdot|\cdot)$ is fixed. We repeat this derivation for our case, where the loss in \eqref{eq:MLLoss} has two parts, representing the contributions of labeled and unlabeled data. Using \eqref{eq:elbo} in the unlabeled data part of \eqref{eq:MLLoss} yields
\begin{align}
	\cL_{\text{ML}} \le
	&-\frac{\gamma}{N_p} \sum_{i=1}^{N_p} \log p_\btheta(s_i, \by_i)
	-\frac{1-\gamma}{N-N_p}\cdot\\
	&\sum_{i=N_p+1}^{N}
	\E_{q(s|\by_i)} \left[-\log q(s|\by_i)+\log p_{\btheta} (s,\by_i)\right]
	\label{eq:EMLoss}
\end{align}
Denote by $\btheta^{(l-1)}$ the estimated parameters vector, $\btheta$, at the beginning of the $l$'th iteration. Then, in the first stage of the $l$'th iteration, we minimize the upper bound on $\cL_{\text{ML}}$, given by the right-hand side of \eqref{eq:EMLoss}, with respect to the conditional distribution, $q(s|\by)$ (while fixing $\btheta=\btheta^{(l-1)}$). By the result mentioned above \cite{bishop2006pattern}, the minimizing conditional distribution is $q(s|\by) = p_{\btheta^{(l-1)}} (s|\by)$.
In the second stage of the $l$'th iteration, we fix $q(s|\by) = p_{\btheta^{(l-1)}} (s|\by)$ in the right-hand side of \eqref{eq:EMLoss}, and minimize the resulting expression  with respect to $\btheta$. Recalling \eqref{eq:p_sy}, this is equivalent to minimizing
$-(\gamma/N_p) \sum_{i=1}^{N_p} \log p_\btheta(\by_i|s_i)
-(1-\gamma)/(N-N_p) \sum_{i=N_p+1}^{N}
\E_{p_{\btheta^{(l-1)}} (s|\by_i)} \left[\log p_{\btheta} (\by_i|s)\right]$
with respect to $\btheta$, which in general is difficult to implement. Instead, we suggest a Monte Carlo EM algorithm (MCEM) \cite{wei1990monte} where, given $\btheta^{(l-1)}$, we sample $s_i^{(l-1)}$ from the distribution $p_{\btheta^{(l-1)}} (\cdot|\by_i)$, for $i = N_p+1,\ldots,N$. Denoting by $l_s(\by) = \log p_\btheta(\by|s)$, given in \eqref{eq:log_pys}, we have by \eqref{eq:p_sy},
\begin{equation}
\label{eq:softmax}
p_{\btheta} (s|\by) = e^{l_s(\by)} / \sum_{j} e^{l_j(\by)}  
\end{equation}
Hence, we can use a softmax transformation to obtain the probabilities of the categorical distribution that we sample from. We note, however, that for each payload data measurement, $y_i$, we need to propagate all possible signals in the constellation through the neural network (fortunately, forward pass only).
Following the sampling process that yields $\left\{ s_i^{(l-1)} \right\}_{i=N_p+1}^N$, we implement a GD iteration on the loss
\begin{align}
	\cL_{\text{MCEM}} =
	&-\frac{\gamma}{N_p} \sum_{i=1}^{N_p} \log p_\btheta(\by_i|s_i)\\
	&-\frac{1-\gamma}{N-N_p}\sum_{i=N_p+1}^{N}
	\log p_{\btheta} (\by_i|s_i^{(l-1)})
	\label{eq:EMLoss2}
\end{align}
Denoting by $\nabla_\btheta \cL_{\text{MCEM}}$, the gradient of $\cL_{\text{MCEM}}$ with respect to $\btheta$, yields the following iterative algorithm, for $l=1,2,\ldots$ (the initial estimate, $\btheta^{(0)}$, is determined by standard neural network parameter initialization),
\begin{equation}
	\btheta^{(l)} = \btheta^{(l-1)} - \eta \cdot \left[ \nabla_\btheta \cL_{\text{MCEM}} \right]_{\btheta=\btheta^{(l-1)}}
	\label{eq:GD_iter}
\end{equation}
where $\eta>0$ is the learning rate (in practice we will use some variant of GD, such as ADAM \cite{kingma2014adam}, applied on mini-batches, see the experiments section). To implement the computation of the gradient in \eqref{eq:GD_iter} efficiently, we apply the back propagation algorithm on the neural network with input $\bx=\bx(s)$, parameters, $\btheta$, and output,
$
\left( \bmu_\btheta(\bx), \log \bsigma_\btheta^2(\bx) \right)
$.

When we apply the GD iterations, $\gamma$ is initially close to 1, in order to assign higher weight to the labeled data, and it monotonically decreases with respect to the iteration index, $l$.

Denoting by $\hat{\btheta}$ the final estimated parameters vector, the decoder uses the following rule to decode the payload data,
\begin{equation}
	\label{eq:decoder_MCEM}
	\hat{s}_i = \argmax_{s\in\{1,2,\ldots,K\}} \log p_{\hat{\btheta}}(\by_i\given s)
\end{equation}
for $i=N_p+1,\ldots,N$, where $\log p_{\btheta}(\by\given s)$ is given by \eqref{eq:log_pys}.

The Viterbi EM decoder is a simpler variant of MCEM. Rather than sampling $s_i^{(l-1)}$ from the distribution $p_{\btheta^{(l-1)}}(\cdot|\by_i)$, we use
$s_i^{(l-1)} = \argmax_{s \in \{1,2,\ldots,K\}} p_{\btheta^{(l-1)}}(s|\by_i)$.
This is a hard decision approach (similarly to SDD), while MCEM uses soft decision.

\subsection{Variational autoencoder (VAE)} \label{sec:VAE}
Reconsider \eqref{eq:EMLoss} and replace $q(s|\by)$ by a parameterized conditional distribution, $q_{\bphi}(s|\by)$. Minimizing the resulting upper bound on $\cL_{\text{ML}}$, yields, due to \eqref{eq:p_sy}, the
loss function,
$-\frac{\gamma}{N_p} \sum_{i=1}^{N_p} \log p_\btheta(\by_i|s_i)
-\frac{1-\gamma}{N-N_p} \cdot \sum_{i=N_p+1}^{N} 
\E_{q_\bphi(s|\by_i)} \left[-\log q_\bphi(s|\by_i)+\log p_{\btheta} (\by_i|s)\right]$,
which is minimized jointly over $\btheta$ and $\bphi$.
Since this loss function trains $\bphi$ using only unlabeled data, we modify it as in \cite{kingma2014semi}:
\begin{align}
\lefteqn{\cL_{\text{VAE}} =
-\frac{\alpha}{N_p} \sum_{i=1}^{N_p} \log q_\bphi(s_i|\by_i) -\frac{\gamma}{N_p} \sum_{i=1}^{N_p} \log p_\btheta(\by_i|s_i) -}\\
&\frac{1-\gamma}{N-N_p} \sum_{i=N_p+1}^{N} 
\left[ \cH\left( q_\bphi(\cdot|\by_i) \right) +
\E_{q_\bphi(s|\by_i)} \log p_{\btheta} (\by_i|s) \right] \quad
\label{eq:VAE}
\end{align}
\begin{equation}
	\cH \left( q_\bphi(\cdot|\by_i) \right) = -\sum_s q_\bphi(s|\by_i) \log q_\bphi(s|\by_i)
	\label{eq:Hq}
\end{equation}
($\cH \left( q_\bphi(\cdot|\by_i) \right)$ is the entropy of the distribution $q_\bphi(\cdot|\by_i)$).
In our experiments, the \emph{VAE-decoder} \cite{kingma2013auto}, $p_\btheta(\by|s)$, is Gaussian, $\cN \left(\by; \bmu_\btheta(\bx(s)), \bsigma_\btheta^2(\bx(s)) \right)$, and a neural network with input $\bx=\bx(s)$ and parameters vector $\btheta$ computes $\left( \bmu_\btheta(\bx), \log \bsigma_\btheta^2(\bx) \right)$, as in Section \ref{sec:MCEM}.
The \emph{VAE-encoder} \cite{kingma2013auto}, $q_\bphi(s|\by)$, is implemented as a neural network with input, $\by$, and a softmax output layer that produces the probability vector, $\left\{ q_\bphi(s | \by) \right\}_{s=1}^K$, as in Section \ref{sec:problem_setup}.

In order to compute the gradient with respect to $\btheta$ of
\begin{equation}
	\E_{q_\bphi(s|\by_i)} \log p_{\btheta} (\by_i|s)
	\label{eq:VAE_middle}
\end{equation}
efficiently, we can approximate it by the unbiased estimator,
$\nabla_\btheta \log p_{\btheta} (\by_i|s_i)$,
where $s_i$ is sampled from the distribution $q_\bphi(\cdot|\by_i)$.
However, it is more difficult to compute the gradient of \eqref{eq:VAE_middle} with respect to $\bphi$, since we would need to apply the back-propagation algorithm on our VAE-encoder neural network $K$ times (once for each value of $s$) for each payload data measurement, $\by_i$, $i=N_p+1,\ldots,N$.
We would also need to apply the forward pass of the VAE-decoder neural network $K$ times for each payload data measurement.
This can be computationally prohibitive for large values of $K$, and intractable for the case of channels with memory.
As an alternative, we apply, similarly to \cite{caciularu2020unsupervised}, categorical reparameterization with Gumbel-softmax \cite{jang2017categorical}, which is a continuous relaxation of the Gumbel-Max trick.
This relaxation enables the use of the reparameterization trick \cite{kingma2013auto} for discrete distributions.
For each $i=N_p+1,\ldots,N$, we first sample $K$ statistically independent uniformly distributed on $[0,1]$ random variables, $u_{i,k}$, $k=1,\ldots,K$, and set
$g_{i,k} = -\log\left(-\log u_{i,k}\right)$
for $k=1,\ldots,K$. The distribution of $g_{i,k}$ is $\text{Gumbel}(0,1)$.
We now define, for $i=N_p+1,\ldots,N$, $k=1,\ldots,K$,
\begin{align}
	\tilde{s}_{i,k} &\defined
	\frac{\exp\left\{ \left( \log q_{\bphi}(k|\by_i) + g_{i,k} \right) / \tau \right\}}
	{\sum_{k'} \exp\left\{ \left( \log q_{\bphi}(k'|\by_i) + g_{i,k'} \right) / \tau \right\}}
	\label{eq:s_ik_sample}\\
	\tilde{\bx}_i &\defined \sum_{k=1}^{K} \bx(k) \tilde{s}_{i,k}
	\label{eq:softmax_app}
\end{align}
(recall that $\bx(k)$ is the $k$'th constellation signal) where $\tau>0$ is a temperature hyper-parameter.
It can be shown \cite{jang2017categorical}, that as $\tau\rightarrow 0$, $\left( \tilde{s}_{i,1},\ldots,\tilde{s}_{i,K} \right)$ will be $\epsilon$-close in $\cL_2$ distance to the $k$'th unit vector with probability $q_\bphi(k|\by_i)$, for $k=1,\ldots,K$, where $\epsilon\rightarrow 0$ as $\tau\rightarrow 0$. As $\tau\rightarrow 0$, $\tilde{\bx}_i$ will be $\epsilon$-close to $\bx(k)$ with probability $q_\bphi(k|\by_i)$, $k=1,\ldots,K$ (with $\epsilon\rightarrow 0$ as $\tau\rightarrow 0$).
Recalling \eqref{eq:log_pys}, our estimate to \eqref{eq:VAE_middle} (up to an additive constant that can be dropped since it does not influence the gradients) is
\begin{align}
	&{\E}_{q_\bphi(s|\by_i)} \log p_{\btheta} (\by_i|s) \approx
	\hat{\E}_{q_\bphi(s|\by_i)} \log p_{\btheta} (\by_i|s)\qquad\\
	&\quad=
	-\frac{1}{2} \sum_{j=1}^{d} 
	\left( \frac{(y_{i,j}-\mu_{\btheta,j}(\tilde{\bx}_i)))^2}{\sigma_{\btheta,j}^2(\tilde{\bx}_i)} + \log \sigma_{\btheta,j}^2(\tilde{\bx}_i)) \right)
	\label{eq:Eqphi}
\end{align}
To summarize, the loss function we use is given by \eqref{eq:VAE}-\eqref{eq:Hq}, where $\E_{q_\bphi(s|\by_i)} \log p_{\btheta} (\by_i|s)$ in \eqref{eq:VAE} is approximated by
$\hat{\E}_{q_\bphi(s|\by_i)} \log p_{\btheta} (\by_i|s)$ in \eqref{eq:Eqphi}.

We set $\tau>0$ to keep $\tilde{\bx}_i$, defined in \eqref{eq:softmax_app}, smooth and differentiable with respect to $\bphi$.
Denoting by $l=1,2,\ldots$ the GD iteration index, we use the following annealing process for $\tau=\tau_l$, the temperature at iteration $l$ \cite{jang2017categorical},
\begin{equation}
	\tau_l = \max(0.5, e^{-0.001 (l-1)} )
	\label{eq:tau_anneal}
\end{equation}
We update $\tau=\tau_l$ every $N_\tau =100$ iterations.

To decode the data, we can use either \eqref{eq:decoder1} (VAE-encoder) or \eqref{eq:decoder_MCEM} (VAE-decoder).
Somewhat better results were obtained by averaging posterior probabilities (i.e., combining
the VAE-encoder and VAE-decoder):
$\hat{s}_i = \argmax_{s\in\{1,2,\ldots,K\}} \left[ 
q_{\hat{\bphi}}(s | \by_i) + p_{\hat{\btheta}}(s | \by_i) \right]$.
Denote $l_s(\by_i) = \log p_{\hat{\btheta}}(\by_i | s)$, which can be computed using \eqref{eq:log_pys}. By \eqref{eq:softmax}, we can compute $p_{\hat{\btheta}}(s | \by_i)$ using a softmax transformation,
while $q_{\hat{\bphi}}(s | \by_i)$ is produced by the VAE-encoder.

\subsection{Comparison with CAVIA meta learning} \label{sec:CAVIA}
Consider a base station that receives block transmissions from devices (users) \cite{park2020learning} in blocks of $N$ symbols, $(s_1,\ldots,s_N)$, over a memoryless channel $p(\by| s)$, as in Section \ref{sec:problem_setup}.
The $m$'th transmitted block of symbols is $\bs^{(m)} = ( s^{(m)}_1,\ldots,s^{(m)}_N )$. The corresponding received channel output at the base station is $\by^{(m)} = ( \by^{(m)}_1,\ldots,\by^{(m)}_N )$.
Each block is transmitted under fixed channel conditions.
We assume that
$\left\{\bs^{(m)}, \by^{(m)} \right\}_{m=1}^M$ are available to the base station (pilot data and / or correctly decoded payload data) when it receives the channel output block corresponding to the test device, $\by^{(0)}$.
We also assume that the first $N_p$ symbols transmitted by the test device, $s_1^{(0)},\ldots,s_{N_p}^{(0)}$, are known pilot symbols. 
Our goal is to estimate the channel based on all the available data at the base station, $\left\{\bs^{(m)}, \by^{(m)} \right\}_{m=1}^M$, $\by^{(0)}$ and $s_1^{(0)},\ldots,s_{N_p}^{(0)}$, and then decode the payload data, $s_{N_p+1}^{(0)},\ldots,s_{N}^{(0)}$.

Various meta-learning methods were examined in \cite{park2020learning} for this problem, including model agnostic meta-learning (MAML) and fast context adaptation via meta-learning (CAVIA) \cite{zintgraf2019fast}.
When only few pilots are available, as in our experiments, CAVIA possesses a lower error rate. Hence, we compare our results to CAVIA.
We denote by $\bz^{(m)}$ a vector that characterizes the channel state for the $m$'th transmitted block, and model by $q_\bphi(s|\by,\bz)$ the conditional distribution of $s=s_i^{(m)}$ given $\by=\by_i^{(m)}$ and $\bz = \bz^{(m)}$. We model this conditional distribution by a neural network with a softmax output layer.
The CAVIA implementation in \cite{park2020learning} estimates $\bphi$ from $\left\{ \by^{(m)}, \bs^{(m)} \right\}_{m=1}^M$. Then, we estimate $\bz^{(0)}$ from $\bphi$, $\by^{(0)}_1,\ldots,\by^{(0)}_N$ and $s^{(0)}_1,\ldots,s^{(0)}_{N_p}$, and decode $s^{(0)}_{N_p+1},\ldots,s^{(0)}_{N}$ with
$q_\bphi ( s_i^{(0)} | \by_i^{(0)}, \bz^{(0)} )$.
We obtained some improvements to this implementation as follows.
Let $p_\btheta\left( \by | s, \bz \right)$ be the conditional PDF of the channel output $\by = \by_i^{(m)}$ given $s=s_i^{(m)}$ and $\bz = \bz^{(m)}$. We use a diagonal covariance Gaussian PDF, $\cN \left( \by;\bmu_\btheta(\bx(s),\bz), \bsigma_\btheta^2(\bx(s),\bz) \right)$ to model this conditional PDF,
\rem{
Denoting by $d$ the dimension of $\by$, we have
\begin{align}
	\lefteqn{\log p_{\btheta}(\by | s, \bz) = -\frac{d}{2} \log (2\pi)}\\
	&\:\: -\frac{1}{2} \sum_{j=1}^{d} 
	\left( \frac{(y_j-\mu_{\btheta,j}(\bx(s),\bz))^2}{\sigma_{\btheta,j}^2(\bx(s),\bz)} + \log \sigma_{\btheta,j}^2(\bx(s),\bz) \right)
	\label{eq:log_pysz}
\end{align}
}
where a neural network with input $(\bx(s),\bz)$, and parameters vector, $\btheta$, produces the mean vector and the vector of log-variances, $\left( \bmu_\btheta(\bx(s),\bz), \log \bsigma_\btheta^2(\bx(s),\bz) \right)$, at its output.
Our modified loss function is based on $\log p_\btheta\left( \by | s, \bz \right) + \log q_\bphi(s|\by,\bz)$ (rather than just $\log q_\bphi(s|\by,\bz)$ in \cite{park2020learning}).

\section{Experiments} \label{sec:experiments}
The setup was described in Section \ref{sec:problem_setup}. As in \cite{park2020learning, cohen2021learning}, consider an internet of things (IoT) system. The transmitted block, $(s_1,\ldots,s_N)$, satisfies $s_i \in \left\{ 1,2,\ldots,K \right\}$ with uniform probability for $K=16$. We use 16-ary quadrature amplitude modulation (16-QAM), i.e., $\bx(s_i) = (x_I(s_i), x_Q(s_i))$ where $x_I(s_i) \in \left\{\pm 1, \pm 3 \right\}$ and $x_Q(s_i) \in \left\{\pm 1, \pm 3 \right\}$. Each symbol, $\bx(s_i)$, experiences I/Q imbalance: The transmitted symbol is $\tilde{\bx}(s_i) = (\tilde{x}_I(s_i), \tilde{x}_Q(s_i))$ where
\begin{equation}
\begin{bmatrix}
	\tilde{x}_I(s_i)\\
	\tilde{x}_Q(s_i)
\end{bmatrix}
=
\begin{bmatrix}
	1+\epsilon & 0\\
	0 & 1-\epsilon
\end{bmatrix}
\begin{bmatrix}
	\cos \delta & -\sin \delta\\
	-\sin \delta & \cos \delta
\end{bmatrix}
\begin{bmatrix}
	{x}_I(s_i)\\
	{x}_Q(s_i)
\end{bmatrix}
\end{equation}
The parameters, $\epsilon$ and $\delta$, are fixed for the entire transmitted block. We set $\epsilon = 0.15 \epsilon_0$ and $\delta = 15\degree \delta_0$, where $\epsilon_0$ and $\delta_0$ are statistically independent random variables, both with a beta distribution, ${\rm Beta}(5,2)$.
The transmitted symbol, $\tilde{\bx}(s_i)$ experiences Rayleigh fading with additive white complex Gaussian noise,
$y_I(s_i) + j y_Q(s_i) = h\left( \tilde{x}_I(s_i) + j \tilde{x}_Q(s_i) \right) + v[i]$,
where $h\sim \cC\cN(0,1)$ (a zero mean, unit variance, complex Gaussian distribution) is fixed over the entire block, and $v[i] \sim \cC\cN(0,\sigma^2)$ are statistically independent. The signal to noise ratio (SNR) is ${\rm SNR} = 10 / \sigma^2$ (since the mean power of $\bx(s_i)$ is $10$).
The output of the channel to the symbol $s_i$ is $\by_i=(y_I(s_i), y_Q(s_i))$ of dimension $d=2$.

We examined the following decoders: Optimal, All Pilots, SDD, MCEM, Viterbi EM, VAE and CAVIA.
The optimal decoder implements ML under the assumption that the channel is known at the receiver. All the other decoders need to estimate the channel.
The All Pilots decoder assumes that all the symbols $\left\{ s_i \right\}_{i=1}^{N}$, are known pilot symbols. The receiver estimates the channel based only on these pilot symbols (standard learning) as was described in Section \ref{sec:problem_setup} using $N_t=5000$ parameter updates.

The SDD, MCEM, Viterbi EM, VAE and CAVIA decoders were described in Section \ref{sec:semisup}. They all use $N_p=16$ pilots.
In the implementation of SDD, MCEM, Viterbi EM and VAE, we use a mini-batch of size $N_b=16$ to process pilot data, and a mini-batch of size $N_u=32$ to process payload data.
The total number of parameter updates was $N_t=5000$.
SDD applies $N_s=1500$ parameter update steps using the loss function \eqref{eq:CE_labeled}, and $N_t-N_s$ parameter update steps using the loss function \eqref{eq:CE_DD}. The best results were obtained for $\gamma_0 = 0.98$.

Rather than using fixed $\gamma$ in \eqref{eq:EMLoss2} and \eqref{eq:VAE}, we used $\gamma_l$ ($l$ is the iteration index), subject to the following annealing process,
$\gamma_l = \frac{1}{1+\beta_l}$, $\beta_l = \min(2 e^{0.0008(l-1)}, \beta_\max)$, where
$\beta_\max = \min ( (N-N_p)/ N_p, 40 )$.
We update $\gamma_l$ and $\beta_l$ every $N_\tau=100$ parameter updates.
According to this scheduling rule, as the iterations proceed, we gradually assign more weight to the unsupervised part of the data.
We also used the annealing schedule described in \eqref{eq:tau_anneal} for the temperature hyper-parameter. We set $\alpha=0.2$ in \eqref{eq:VAE}.

The values of the parameters used are summarized in Table \ref{tb:parameters}.
We used ADAM with learning rate $0.001$.
In the implementation of CAVIA we used $M=1000$, $N=3200$ and ${\rm dim}(\bz)=10$.
\begin{table}[!htbp]
	\begin{center}
		\caption{Values of parameters used in our experiments}
		\label{tb:parameters}
		\begin{tabular}{|l|l|l|l|l|l|l|l|l|l|}
			\hline
			$K$  & $d$ & $N_p$ & $N_b$ & $N_u$ & $N_t$  & $N_s$  & $\gamma_0$ & $N_\tau$ & $\alpha$ \\ \hline
			$16$ & $2$ & $16$  & $16$  & $32$  & $5000$ & $1500$ & $0.98$      & $100$    & $0.2$ \\ \hline
		\end{tabular}\\
	\end{center}
\end{table}
\rem{
\begin{table}[!htbp]
	\vspace{4mm}
	\begin{center}
		\caption{Values of parameters specific to CAVIA}
		\label{tb:parameters_CAVIA}
		\begin{tabular}{|l|l|l|l|l|}
			\hline
			$M$     & $N$     & $|\cM'|$ &  {\rm dim}(\bz) \\ \hline
			$1000$  & $3200$  & $40$     &  $10$  \\ \hline
		\end{tabular}\\
	\end{center}
\end{table}
}

\rem{
On the other hand, CAVIA uses stochastic GD (SGD) with momentum $0.9$ rather than ADAM since it performed better. The learning rate was $0.2$ both for the update with respect to $\bz^{(m)}$ during meta-training, and for the first $L_u$ updates with respect to $\bz^{(0)}$ during meta-testing. The learning rate for all the other updates was $0.01$.
In the CAVIA experiments we have also applied exponential decay scheduling for the learning rate of $\bphi$ and $\btheta$: Every $100$ training iterations we multiply the learning rate by $0.93$ (the initial learning rate is $0.01$).
}

Shallow neural networks with small number of parameters were sufficient for obtaining good results.
We used four neural networks in our Pytorch implementation (a GPU was not required). The first two for implementing $q_{\bphi}(s|\by)$ and $q_{\bphi}(s|\by,\bz)$ (CAVIA experiment), both with a softmax output layer of size $K=16$. The other two are for implementing the 
mapping from $\bx=\bx(s)$ to $(\bmu_\btheta(\bx), \log \bsigma^2_{\btheta}(\bx))$ and the mapping from $(\bx,\bz)$ to $(\bmu_\btheta(\bx,\bz), \log \bsigma^2_{\btheta}(\bx,\bz))$ (CAVIA), both with an output layer of size $4$.
All neural networks had three fully connected hidden layers of sizes $10$, $30$ and $30$.

\begin{figure}[!htbp]
\includegraphics[width=\linewidth]{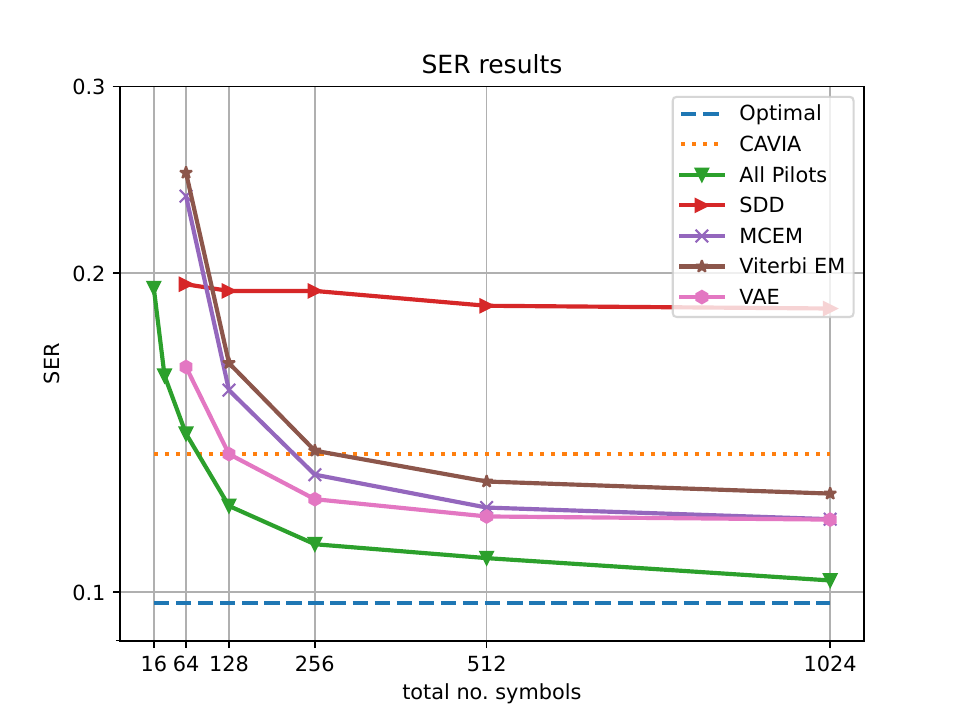}
\caption{SER results as function of the total number of symbols, $N$, for SNR=$18\text{dB}$. In CAVIA, SDD, MCEM, Viterbi EM and VAE, the number of pilot symbols was $N_p=16$.}
\label{fig:SER_vs_SNR_18}
\end{figure}

\begin{figure}[!htbp]
\includegraphics[width=\linewidth]{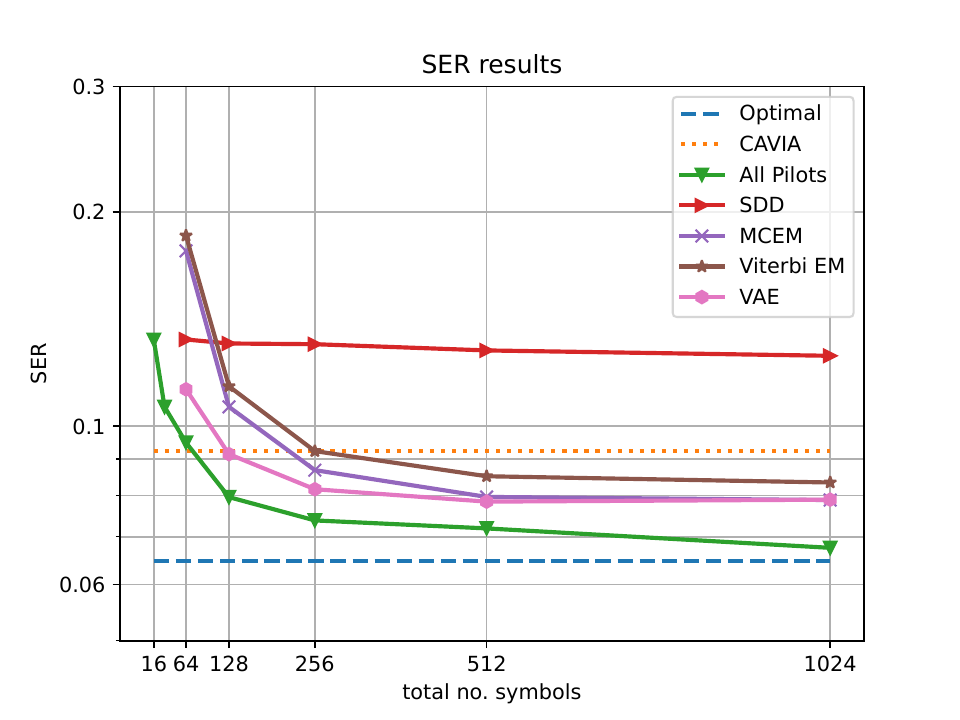}
\caption{Same as Fig. \ref{fig:SER_vs_SNR_18} for SNR=$20\text{dB}$.}
\label{fig:SER_vs_SNR_20}
\end{figure}

In Figs. \ref{fig:SER_vs_SNR_18} and \ref{fig:SER_vs_SNR_20} we plot the symbol error rate (SER) of the various decoders for SNR=$18\text{dB}$ and SNR=$20\text{dB}$.
Each SER result was obtained by averaging over $2000$ test devices, each with $1000$ symbols.
In addition to that, the experiments with CAVIA meta learning were repeated for $10$ different training sessions, and the average was taken.
\rem{
When using the original CAVIA version \cite{park2020learning} based on $q_\bphi(s|\by,\bz)$ only (not shown in the figures), for SNR=$20\text{dB}$, one out of $10$ CAVIA train sessions did not converge to a low SER value. The other $9$ had an average SER of $0.09731$. On the other hand, with the modified CAVIA described in Section \ref{sec:CAVIA}, all $10$ CAVIA train sessions were successful, and their average SER value was $0.09225$ as reported in Fig. \ref{fig:SER_vs_SNR_20}.
}

It can be seen that MCEM, Viterbi EM and VAE perform much better than SDD. The VAE outperforms the other SSL methods. It saturates at about 512 symbols (16 pilot symbols and 496 payload symbols).
When the number of payload symbols is larger than $112$, the SER of VAE is lower than that of CAVIA with the same number of pilot symbols, $N_p=16$. This may be explained by the inductive bias inferred by meta learning \cite{park2020learning}. Note that meta learning assumes the availability of pilots or correctly decoded data from previous transmissions to the base station, under different channel realizations.

\section{Conclusion}
We presented SSL methods for variational inference over nonlinear channels, and demonstrated the best results for a VAE decoder that exploits neural networks. 
For the considered communication example, with sufficiently many payload symbols, SSL with a VAE outperformed meta learning. The VAE is simple to implement, and it does not require the additional information and resources that meta learning assumes. It is robust, since it does not assume that the current and previous channel realizations are members of the same family. 
Instead, it relies only on the current information block (including pilot symbols) that needs to be decoded. 
The extension to SSL over nonlinear channels with memory is left for future research.



\end{document}